\documentclass[superscriptaddress,
reprint,nofootinbib,
preprintnumbers,
amsmath,amssymb,aps,prl]{revtex4-1}

\pdfoutput=1
\usepackage{balance}
\usepackage[utf8]{inputenc} 
\usepackage[T1]{fontenc} 
\usepackage{microtype}
\usepackage{amsmath,amssymb}
\usepackage{verbatim}
\usepackage{amsmath,appendix}
\usepackage{amsthm}
\usepackage{mathtools}

\usepackage{graphicx}
\usepackage[dvipsnames]{xcolor}

\usepackage{dcolumn}
\usepackage{bm}

\usepackage[colorlinks = true,
            linkcolor = blue,
            urlcolor  = blue,
            citecolor =red,
            anchorcolor = blue]{hyperref}
\usepackage{verbatim}
\usepackage{color,ulem}
\usepackage[english]{babel}

\usepackage{blindtext}
\usepackage{float,amsmath,amssymb}

\newcommand{\beq}{\begin{eqnarray}}
\newcommand{\eeq}{\end{eqnarray}}

\newtheorem{theorem}{Theorem}[section]
\usepackage{graphicx}
\usepackage{amsmath}
\usepackage{gensymb}
\usepackage{physics}
\usepackage{balance}
\usepackage[varg]{txfonts}
\usepackage{ulem,lipsum}
\usepackage{fancyhdr}

\begin{document}

\preprint{IFT-UAM/CSIC-20-82}
\title{\huge Anomalous diffusion and Noether’s second theorem\color{black}}
\author{Matteo Baggioli}
\email{matteo.baggioli@uam.es}
\affiliation{Instituto de Fisica Teorica UAM/CSIC, c/ Nicolas Cabrera 13-15, Cantoblanco, 28049 Madrid, Spain}
\author{Gabriele La Nave}
\email{lanave@illinois.edu }
\affiliation{Department of Mathematics, University of Illinois, Urbana, Il. 61801}
\author{Philip W. Phillips}
\email{dimer@illinois.edu }
\affiliation{Department of Physics and Institute for Condensed Matter Theory, University of Illinois
1110 W. Green Street, Urbana, IL 61801, U.S.A.}

\begin{abstract}
Despite the fact that conserved currents have dimensions that are determined solely by dimensional analysis (and hence no anomalous dimensions), Nature abounds in examples of anomalous diffusion~\cite{cells,anomdiffexpt} in which $L\propto t^\gamma$, with $\gamma\ne 1/2$, and heat transport in which the thermal conductivity diverges as $L^\alpha$~\cite{mcuentherm,therm2,therm3,therm4,therm5}.  Aside from breaking of Lorentz invariance, the true common link in such problems is an anomalous dimension for the underlying conserved current, thereby violating the basic tenet of field theory.  We show here that the phenomenological~\cite{klafter,fracdiff2016,nigmatullin} non-local equations of motion that are used to describe such anomalies all follow from Lorentz-violating gauge transformations arising from N\"other's second theorem.  The generalizations lead to a family of  diffusion and heat transport equations that systematize how non-local gauge transformations generate more general forms of Fick's and Fourier's laws for diffusive and heat transport, respectively. In particular, the associated Goldstone modes of the form $\omega\propto k^\alpha$, $\alpha\in \mathbb R$ are direct consequences of fractional equations of motion.
\end{abstract}

\maketitle
\section{Introduction}

Although Lorentz symmetry is a fundamental organizing principle of the universe, it is oftentimes violated.  For example, any collection of particles, complicated enough for a center of mass to be well defined, has a preferred reference frame and hence violates the boost-invariance of the Lorentz group.  The lack of boost-invariance rarely occurs as a solo act. The presence of a lattice and non-relativistic dispersions, graphene the exception, in condensed matter systems offer additional routes for the breakdown of Lorentz invariance. It comes as no surprise that low-energy phases of matter can be indeed classified \cite{Nicolis:2015sra} by looking at the different spontaneous symmetry breaking patterns of the Lorentz group. While continuous symmetry breaking such as rotational and translational give rise to Goldstone bosons of the form, $\omega\propto k^n$, where $n\in \mathbb Z$, the massless excitation arising from just the breaking of boost-invariance remains elusive.  Of course, such a problem would be rendered moot if phases of matter exist in which just \cite{Nicolis} boost-invariance were broken.  Nature, unfortunately is not so kind.

Our focus here is on two processes that break Lorentz invariance, diffusion and heat transport.  The equations for both are derivable from the continuity equation,
\begin{equation}
    \partial_\mu J^\mu=\partial_t\,n\,+\,\partial_i J^i=0\label{continuity}
\end{equation} 
which arises from an integration by parts of the action
\beq
S=\int d^dx\, \left(F^2+J^\mu A_\mu\right)
\eeq
subject to the $U(1)$ gauge-invariant condition, $A_\mu\rightarrow A_\mu-\partial_\mu\Lambda$ with $F=dA$.  For simple diffusion, the inconsistency with Lorentz invariance occurs once Fick's law for the current,
\beq
J^i\,=\,-\,D\,\partial^i n \label{fick}
\eeq
is imposed.  Substituting \eqref{fick} into the continuity equation \eqref{continuity} results in the well known diffusion equation
\beq
\partial_t n - \,D\,\Delta\,n=0.
\eeq
The propagator for this equation has solutions even for space-like separations.   Such growth is inconsistent with special relativity and hence must be associated with a massless mode, in this case $\omega=-iDk^2$.  Arguments of this sort have motivated an extensive literature~\cite{goldstein,dudley,goldstein2} on squaring standard diffusion with special relativity.  A promising solution~\cite{hanggi} appears to be random walks based on non-Markovian processes, as relativistic Markov dynamics in space is a misnomer.  The inconsistency of diffusion with special relativity is also known in the context of linearized relativistic hydrodynamics where it manifests itself as the violation of causality -- superluminal propagation. A famous resolution to this problem was given by Israel and Stewart \cite{ISRAEL1979341} and it can be reformulated as an upper bound for the diffusion constant $D<v^2 \tau$ \cite{PhysRevLett.119.141601}, where $v$ is the lightcone speed and $\tau$ the equilibration time at which diffusion obtains.

What the previous derivation lays plain is that the form of the current imposed by Fick's law fixes the diffusion equation.  However, Fick's choice is not unique and, as a result, neither is the diffusion equation. In fact, there is an extensive literature, on replacing either the time derivative~\cite{cafffractime,nigmatullin} or the spatial Laplacian~\cite{klafter,zaslavsky,fracdiffcrit,fracfickslaw,caffdiff,fracdiff2016} in the diffusion equation with a pseudo-differential operator such as the fractional Laplacian.  What results from such analyses are non-local diffusive transport or L\'evy processes that are capable of describing the numerous experimental realizations of  anomalous diffusion~\cite{anomdiffexpt,cells,expt2,ramos} in which $L\propto t^\delta$,  $\delta\ne 1/2$.  \textit{Subdiffusion} corresponds to $\delta<1/2$, whereas \textit{superdiffusion} arises when $\delta>1/2$. 

Anomalies also occur for heat transport which is derivable immediately from the continuity equation by substituting Fourier's law for the gradient of the temperature profile
\beq
J_H=-\kappa \,\frac{\partial T(x,t)}{\partial x},
\label{fourier}
\eeq
where $\kappa$ is the thermal conductivity. From dimensional analysis $J_H\propto L^{-1}$ and $\kappa$ scales as $L^0$.   However, since the pioneering paper of Fermi, Pasta and Ulam~\cite{fpu} showed that even in an anisotropic harmonic solid $\kappa$ diverges, numerous physical systems~\cite{dhar,povstenko1,Lepridivtc,heatcond2} as simple as single molecules~\cite{therm5} and nanotubes~\cite{mcuentherm,zettl,therm2,therm3,therm4} have been shown to violate Fourier's law as they exhibit a thermal conductivity diverging as $\kappa\propto L^\alpha$. Such a divergence of $\kappa$ implies that the heat current scales as $J_H\propto L^{\alpha-1}$. Only $\alpha=0$ corresponds to standard diffusive transport.  As in anomalous diffusion, fractional equations of motion~\cite{tf1,tf2,tf3} are also invoked to explain a heat current that scales with an anomalous power, $J_H\propto L^{\alpha-1}$. What is surprising with the heat current is that the scaling $\kappa^\alpha$ applies~~\cite{garrido} even to the archetypeal 1d diatomic hard-point gas implying that a fundamental non-locality underlies the heat transport even in this simple case.  From the Weidemann-Franz law, any anomaly in the heat transport is passed onto the electrical transport.  Equivalently, heat transport in charged fluids is governed by $U(1)$ invariance and the derivation presented in the first section applies.  The standard folklore~\cite{gross} is that any current tied to a $U(1)$ gauge field cannot acquire an anomalous dimension under any amount of renormalization.  Hence, squaring any of the non-local schemes~\cite{dhar,tf1,tf2,tf3} that have been devised to treat anomalous heat transport, that is $\kappa\propto L^{\alpha}$~\cite{garrido,zettl,mcuentherm,zettl,therm2,therm3,therm4} or anomalous diffusion~\cite{tf1,tf2,tf3,klafter,zaslavsky,fracdiffcrit,fracfickslaw,caffdiff} must account for how they get around the standard argument that the dimension of the conserved current is fixed solely by dimensional analysis.  

It is precisely this problem that we address in this paper.  In previous work, we have shown~\cite{csforms,gl1} that N\"other's second theorem (NST) allows for non-traditional scaling of the current only if the action of the gauge group is changed fundamentally. However, all of our extensions of N\"other's Second theorem presented previously~\cite{csforms,gl1} preserve Lorentz invariance and hence are not directly applicable to anomalous heat or diffusive transport.  What we do here is generalize these arguments to the non-Lorentzian case. We focus on the minimal constraints that must be placed on the action of the gauge group to derive anomalous diffusive transport equations. We show that the phenomenological diffusion equations~\cite{tf1,tf2,tf3,klafter,zaslavsky,fracdiffcrit,fracfickslaw,caffdiff} have a rigorous basis in N\"other's second theorem.   This theorem stems from the simple observation that the continuity equation $\partial_\mu J^\mu=0$ is unique up to any operator, $\hat Y$ that commutes with the exterior derivative resulting in
\beq
    \partial_\mu \hat{Y}J^\mu=\partial_\mu \tilde{J}^\mu=0
\eeq
being an equally valid equation of motion (EOM).  To be consistent, the operator $\hat Y$ must also be present in the gauge-invariant condition.  Stated succinctly, the second theorem finds that the full family of generators of $U(1)$ invariance determines the dimension of the current not just the linear derivative term used earlier.  In general, the second theorem applies anytime there are either a collection of infinitesimal symmetries or one symmetry parameterized by an arbitrary number of functions. More recent examples that fall into this category are \textit{fractons} \cite{Gromov:2020yoc}, exotic excitations with confined mobility \cite{doi:10.1142/S0217751X20300033}.  Such anomalous behaviour was  connected to the conservation of higher-order \textit{charges} such as the dipole moment and to new tensorial symmetries \cite{PhysRevB.95.115139,Gromov:2018nbv,Pretko:2018jbi,Seiberg:2020bhn} and follow necessarily from NST.

We have exploited the degeneracy implied by N\"other's second theorem to show~\cite{csforms,gl1,varna} that the most general formulation of electricity and magnetism is one with fractional equations of motion with full Lorentz symmetry.  However, to describe diffusion, we need to restrict ourselves solely to a subset of operators that ultimately breaks Lorentz invariance explicitly. From our formulation, we see that the associated Goldstone is now $\omega\propto k^\alpha$ with $\alpha\in \mathbb R$.  Goldstone modes with fractional powers have been observed previously as ripplon modes in a 2D Wigner crystal~\cite{grimes,halperin}, which have the same dispersion as plasmons in a 2D electron gas, and domain-wall fluctuations at superfluid-superfluid interfaces~\cite{mazets} as they obey $k^{1/2}$ and $k^{3/2}$ dispersions, respectively.  In the context of the latter, a non-local action~\cite{haruki}  containing the fractional Laplacian acting on the displacement field for the interface was proposed to generate the corresponding gapless boson dispersion. The appearance of the fractional Laplacian was justified based on an integration of gapless degrees of freedom~\cite{haruki}.   We show here that such fractional dispersions result quite generally from N\"other's Second theorem.\\
The results presented in this work are very general and applicable to any system exhibiting a macroscopic anomalous diffusive dynamics, from soft-matter systems and biological structures (intracellular transport \cite{santamaria2009protein}, inhomogeneous media \cite{durang2015overdamped}, DNA polymer chains \cite{lomholt2005optimal}, monolayers of motile cells \cite{palmieri2015multiple}) to complex fluids, hard matter and cosmological evolution (protostellar birth \cite{vaytet2018protostellar}, viscoelastic systems \cite{mason1995optical}, electron-hole plasma in semiconductor quantum wells \cite{borges2006optical}, heterogeneous structures \cite{filipovitch2016infiltration}). Our framework, entirely based on symmetry principles, encompasses all these situations in a unified picture.\color{black}

\section{Background}
 
The power of NST in gauge theories can be understood from a simple degeneracy argument.  Consider the Maxwell action,
\beq\label{Maction}
S&=&\frac12\int \frac{d^dk}{2\pi^d}\, A_\mu(k)\,\left[k^2 \eta^{\mu\nu}-k^\mu k^\nu\right]\,A_\nu(k)\nonumber\\
&=&\frac12\int d^dk\, A_\mu(k)\,M^{\mu\nu}\,A_\nu(k).
\eeq
All gauge transformations  appear as zero-eigenvalues of $M$.  For example,
\beq
M^{\mu\nu}k_\nu=0,
\eeq
which yields the standard gauge-invariant condition in electromagnetism because  $ik_\nu$ is just the Fourier transform of $\partial_\nu $.
The ambiguity that leads to NST comes from noticing that if $k_\nu$ is an eigenvector, then so is $f(k) k_\nu$, where $f(k)$ is a scalar which depends on the momentum itself. Whence, there are a whole family of eigenvectors,
\beq\label{mgen}
M_{\mu\nu}\,f(k)\, k^\nu=0,
\eeq
that satisfy the zero eigenvalue condition, each characterizing a perfectly valid electromagnetism (EM).   It is for this reason that N\"other~\cite{nother} devoted the second half of her paper to the consequences retaining all possible integer derivatives,
\beq
\label{Nothergen}
A_\mu\rightarrow A_\mu +\partial_\mu\Lambda +\partial_\mu\partial_\nu G^\nu+\cdots,
\eeq
 in the gauge-invariant condition for $A_\mu$ has on the conservation laws for the current. All higher-order integer derivatives generate new constraints on the current and correspond to new conserved conserved quantities, dipoles for example \cite{PhysRevB.95.115139,Gromov:2018nbv,Pretko:2018jbi,Seiberg:2020bhn}.
To determine $f$, we note the following. We now take $\bf p$ to the the full 4-momentum.  1) $f$ must be rotationally and Lorentz-invariant.  2) $f$ cannot change the fact that $\Lambda$ is dimensionless; equivalently it cannot change the fact that $A$ is a 1-form.  3) $f$ must commute with the total exterior derivative; that is, $[f,p_\mu]=0$ just as $[d,{\hat Y}]=0$.  Hence, finding $f$ is equivalent to fixing $\hat Y$.  A form of $f$ that satisfies all of these constraints is $f\equiv f(p^2)$.   In momentum space, $k^2$ is simply the Fourier transform of the Laplacian, $-\Delta$.  As a result, the general form of $f(p^2)$ in real space is just the Box operator
raised to an arbitrary power, and the generalization in Eq. (\ref{mgen}) implies that there are a multitude of possible electromagnetisms  (in vacuum) that are
 invariant under the transformation,
 \beq
 A_\mu\rightarrow A_\mu + f(p^2)\,i\,p_\mu\,\Lambda,
 \eeq
 or in real space,
  \beq
  A_\mu\rightarrow A_\mu + (-\Box)^\gamma \partial_\mu\Lambda,
  \eeq
  resulting in $[A_\mu]=1+2[f]=\gamma$.  The definition of the fractional Laplacian and fractional Box operator we adopt here is determined by the general form of a differential operator $L$
  \beq\label{reisz2}
 (L_x)^\gamma f(x)= \frac{1}{\Gamma(\gamma)}\int_0^{+\infty} e^{-t\,L} f(t)\, \frac{dt}{t^{1-\gamma}}
 \eeq
 where $e^{-tL}$ is the "heat" flow of the operator $L$, i.e. $\beta :=e^{-tL}f$ is the unique solution to $\partial _t \beta + L\beta =0$ with Dirichlet boundary condition $\beta\mid_{t=0}= f$.
In the case of the Laplacian acting on functions, one can prove this coincides with the standard Riesz definition
\beq\label{reisz}
(-\Delta_x)^\gamma f(x)=C_{n,\gamma}\int_{\mathbb R^n}\frac{f(x)-f(\xi)}{\mid {x-\xi}\mid^{n+2\gamma}}\;d\xi
\eeq
for some constant $C_{n, \gamma}$.  Note, rather than just depending on the information of $f(x)$ at a point, the fractional Laplacian requires information everywhere in ${\mathbb R}^n$.  For the Laplacian acting on differential forms, there is a similar formula which holds componentwise, as it was shown by the authors in \cite{csforms}.
The standard Maxwell theory is just a special case in which $\gamma=1$.  In general, the theories that result for $\gamma\ne 1$ allow for the current to have an arbitrary dimension, not necessarily $d-1$, without spoiling its conservation nor violating Gross' argument \cite{gross}.  Identifying $\hat Y$ with the fractional Box operator yields the conservation law
\beq \label{fractchargecons}
\partial^\mu (-\Box)^{(\gamma-1)/2}J_\mu=0.
\eeq
\noindent
 As expected, this ambiguity shows up at the level of the Ward identities.  The current-current correlator for the photon
\beq
C^{ij}(k)  \propto (k^2)^{\gamma}\, \bigg( \eta^{ij} - \frac{k^ik^j}{k^2} \bigg)
\eeq
does not just satisfy  $k_\mu C^{\mu\nu}=0$ but also $k^{\gamma-1} k_\mu C^{\mu\nu}=0$. This translates into either $\partial_\mu C^{\mu\nu}=0$, the standard Ward identity, or
 \beq
 \partial_\mu (-\Box)^{\frac{\gamma-1}{2}}C^{\mu\nu}=0
 \eeq
which illustrates that the current conservation equation only specifies the current up to any operator that commutes with the total differential.    Conservation laws such as the one in Eq. \eqref{fractchargecons} are in some sense more fundamental, as one can infer the standard ones from them, but more importantly they can occur earlier~\cite{gl1,csforms} in the hierarchy of conservation laws that stem from N\"other's first theorem.\\

\section{Non-Lorentz Invariant Theory}

We start by discussing the form of the effective Lagrangian of a non-boost invariant theory (which may or may not preserve Galilean transformations). Such an effective Lagrangian can be written as
\beq
L= \frac{1}{2}\int \, d^d x \left( {\mathcal K}F_{ij}\, {\mathcal K}F^{i j}+ c\, F_{0i}^2\right) + L_m,
\eeq
where ${\mathcal K}F_{ij}= \int \,  K(x,y) \left( F_{ij}(x) -F_{ij}(y)\right)d^dy$ with $K(x,y)$ a function of the ~2 vector variable $x$ and $y$ (and we assume, naturally, that $K(x,y)=f(|x-y|)$ in case the Galilean symmetry is not broken and there is no loss of generality in assuming that $c$ is constant. For simplicity of calculations in recovering our eventual equations of motion, one should take $f(u)=u^{-n+1-\gamma},$ so that ${\mathcal K} F=(-\Delta)^{\frac{\gamma-1}{2}}$ to be consistent with Eq. (\ref{reisz}) and $d=n+1$. The matter-field Lagrangian $L_m$ also has to be of the form 
\begin{widetext}
\begin{equation}
\resizebox{\hsize}{!}{$L_m= \int \, \left( \int \left(c_1(s,t) (\partial_t \phi)^2(x,s)\right) ds \right) d^dxdt+\int \left( \int \left(\,-\,c_2(x,y)\,(\partial_i \partial_j \phi)^2(y)-c_3(x,y) (\nabla^2 \phi)^2(y)+\dots \right)d^dy\right) d^dxdt+F_1(\nabla A-\phi)+F_2(A_0-\phi_t)$},
\end{equation}
\end{widetext}
and again without loss of generality, we can assume $\int \left(c_1(s,t) (\partial_t \phi)^2(y)\right) ds= (\partial_t \phi)^2$ and that in the case that Galilean symmetries are preserved, $c_k(x,y)=f_k(|x-y|^2).$
Further, the condition that boost symmetry be broken, is equivalent to requiring that the {\it pseudo-differential} operators $\int \left(c_1(s,t) (\partial_t \phi)^2(x,s)\right) ds$ and $ \int \left(\,-\,c_2(x,y)\,(\partial_i \partial_j \phi)^2(y)-c_3(x,y) (\nabla^2 \phi)^2(y)+\dots \right)d^dy$ be of different {\it order}.

Although most of our discussions apply to the general setting, nonetheless, for the sake of clarity, we will focus our attention on the case that the pseudo-differential operators are fractional Laplacians.
To apply this framework to diffusion, we break the Lorentz-invariant condition on $f(p^2)$.  
In order to define the corresponding Lagrangian, we discuss a more general setting at first. We thus concern ourselves with the context in which we are given a mixed tensor $\Omega$ which for clarity of exposition we think of as being represented by a pair of tensors $A$ and $B$. In fact, since we are concerned with EM (or Yang-Mills theory more generally) we think of our mixed tensor $\Omega$ as being the sum of $2$ differential forms $A$ and $B$. We assume that the underlying manifold separates as $M=M_1\times M_2$. In this application, such a decomposition arises from the the switch from the Einsteinian to the Galilean perspective as a result of symmetry breaking.  Correspondingly, $M= \mathbb R\times M_2$ in this case (in particular this occurs for globally hyperbolic spacetimes).\begin{footnote}{ We don't really need the manifold to split as a product in the discussion that follows, but we just need the existence of a map $\pi: M\to M_2$ and a {\it horizontal} distribution of $TM$ relative to $\pi$.}\end{footnote}
This results in a decomposition on $r$-forms $\Omega ^r (M)=\bigoplus _{p+q=r}\, \Omega ^p  (M_1) \otimes \Omega ^q (M_2).$ Our notation here corresponds to regarding $ \pi_i^* \Omega ^s  (M_i) $ (for $i=1,2$) with $\Omega ^s  (M_i) $; that is an element of $\Omega ^s  (M_i) $ is understood as a linear combination $\alpha _{\ell _1 , \cdots \ell _s} dx^{\ell _1} \wedge \cdots \wedge dx^{\ell_s}$, where the $ \alpha _{\ell _1 , \cdots \ell _s} $'s are functions on the entirety of $M$ and $dx^{\ell _1},\cdots, dx^{\ell_s}$ arise from coordinates on $M_i$ only.
Letting $A$ and $B$ be represented in  coordinates as $A=A_I dx^I$ and $B=B_Jdx^J$, we use a multi-index notation, where $I=(i_1,\cdots , i_p)$ and $J=(j_1,\cdots , j_q)$, so $A= A_{i_1,\cdots, i_p} dx^{i_1}\wedge \cdots dx^{i_p}$, etc. The work horse of the non-Lorentzian picture will be the operator $D_\gamma$,
\begin{equation}
D_\gamma \Omega= d_1 \Omega +d_2\Delta _2 ^{\frac{\gamma -1}{2}} \Omega=  d_1 A+d_2\Delta _2 ^{\frac{\gamma -1}{2}} A+  d_1 B +d_2\Delta _2 ^{\frac{\gamma -1}{2}} B,
\end{equation}
where $d_i$ is the differential coming from $M_i$ and $\Delta _2= d_2d_2^{*_2}+ d_2^{*_2}d_2$ is the Hodge Laplacian on forms, where $*_2$ is the Hodge star operator on $\Omega ^r_2$  . This is clearly linear with respect to the decomposition of $\Omega$ when $\Omega= A+B$.  Thus we analyse it in terms of its coordinates for a given form. Let us indicate coordinates $x^i$ according to a separation arising from the split $M=M_1\times M_2$. So we write the coordinates $\{x^i\}_{i=1}^n$ for those coming from $M_1$ and $\{x^i\}_{i=n+1}^d$ for the ones coming from $M_2$. Whence, in terms of coordinates, for $A$ (and analogously for $B$), we have
\begin{equation} \begin{aligned} &D_\gamma A = \sum _{i+1}^n \frac{\partial } {\partial x^{i}} A_I\, d x^i\wedge dx^I+\sum _{i=n+1} ^d \frac{\partial } {\partial x^{\mu_i}}\left( \Delta _2^{\frac{\gamma-1}{2}}A_I\right) d x^{\mu_i}\wedge dx^I.\end{aligned}\end{equation}
In the special case (that we care about here) of breaking boost symmetry,
\begin{equation} \begin{aligned} &D_\gamma A = \frac{\partial } {\partial x^{0}}\, A_I \,d x^{0} \wedge dx^I+\sum _{i=1} ^d \frac{\partial } {\partial x^i}\left( \Delta _x^{\frac{\gamma-1}{2}}A_I\right) d x^i\wedge dx^I,\end{aligned}\end{equation}
where $\Delta _x^{\frac{\gamma-1}{2}}A_I$ is the fractional Laplacian on the spatial coordinates. In other words, 
\begin{equation}
D_\gamma A = d_1 A+ d_{2,\gamma}\,A
\end{equation}
where $d_{2,\gamma}= d_2 \Delta _2^{\frac{\gamma-1}{2}} A$ is the fractional differential introduced in \cite{csforms} and can be computed using Eq (\ref{reisz}).

If we denote by $\star$ the usual Hodge Laplacian associated with a product metric $ds^2= ds_1^2 + ds_2^2$, it is a straightforward matter to show that 
\beq
D_\gamma^*:= \star D_\gamma \star
\eeq
is an operator that takes a $p$-form to a $(p-1)$-form and it is the (formal) adjoint of $D_\gamma$ with respect to the standard scalar product on forms. We can show
\begin{theorem}
For any $p$-form $A = A_I \,dx^I$ we have
\begin{equation}
(D_\gamma ^* D_\gamma + D_\gamma D_\gamma ^* ) \,A= (\Delta _1 A_I )\,d^I + (\Delta _2 ^\gamma A_I)\, dx^I
\end{equation}
and in Lorentzian signature 
\begin{equation}
(D_\gamma ^* D_\gamma + D_\gamma D_\gamma ^* ) \,A= (\Delta _1 A_I )\,d^I - (\Delta _2 ^\gamma A_I)\, dx^I
\end{equation}
In particular if we start with the mixed form $A=A_0\, dt+ A_{ij}\,dx^i\wedge dx^j$
\begin{equation}
\resizebox{\hsize}{!}{$(D_\gamma ^* D_\gamma + D_\gamma D_\gamma ^* ) A= \left( \frac{\partial ^2A_{0}}{\partial t^2}-\Delta_x ^\gamma A_ 0 \right)  dt  +\left( \frac{\partial ^2A_{ij}}{\partial t^2}  -\Delta _x^\gamma A_{ij}\right)dx^i \wedge dx^j\nonumber$}
\end{equation}
\end{theorem}

We now define a non-local $U(1)$-Gauge theory for a multi-form $\Omega$
\beq\label{non-lorentzEM} 
D_\gamma \left( \star D_\gamma \Omega\right)=\star J,
\eeq
where $J$ is a mixed-form. We will specialize to the case that $\Omega = A_0 dt+ A_{ij} dx^i dx^j$ and then $J$ will be the form dual to the current $(d+1)$-vector $(\rho,J ^{ij})$. Equation \eqref{non-lorentzEM} is readily seen, after determining the Gauge fixing condition $D_\gamma ^* \Omega=0$, to be equivalent to 
\beq 
(D_\gamma ^* D_\gamma + D_\gamma D_\gamma ^* )\, \Omega= \star J.
\eeq
Writing as usual $ A_{ij}$ for the vector corresponding to the spatial components of $A$, Eq. \eqref{non-lorentzEM} is thus seen to be equivalent to 
\begin{align} &\left( \frac{\partial ^2}{\partial t^2}-\Delta ^\gamma\right)  A_0 = \rho \\&  \left( \frac{\partial ^2}{\partial t^2}  -\Delta ^\gamma \right)\vec {A}= \vec {J}
\end{align}
the non-Lorentzian equations of motion for the conserved currents.

 The action of the Gauge group that yields such currents is easy to express in the case that $A$ is a Lie-valued $1$-form. We briefly described this without proofs since it follows from a simple generalization of what the authors did in \cite{csforms}. We describe it in the $U(1)$-case only. First, we note that the $U(1)$-action is determined by a diagonal action on a multi-form $\Omega=A_0+A_1$ (here $A_0$ and $A_1$ are both $1$-forms, so the decomposition is only meant to evoke the decomposition $M=M_1\times M_2$), meaning that given an element $e^{i\lambda}\in U(1)$, it will act on the individual components of $\Omega$ separately. We establish the following fact about our $U(1)$-Gauge action, denoting by $\Box_{(s_1,s_2)}:= \Delta _1^{s_1}-\Delta _2^{s_2}$ and in fact, as we argued, the most significant case is when $s_1=1$ and $s_2$ is arbitrary (as we can always reduce ourselves to this up to fields redefinitions). In what follows, we then write $\Box_s$ for $\Box_{1,s}.$ 
\begin{theorem}
We define an action of $U(1)$ on {\it local} sections of the principal bundle corresponding to $A$ as
\beq
e^{i\Lambda} \odot \phi= \Box_ \frac{1-\gamma}{2}\left( e^{i \Box_ \frac{\gamma-1}{2} \Lambda} \Box_ \frac{\gamma-1}{2}\phi\right),
\eeq
and we define the non-Lorentz invariant covariant derivative as
\beq 
\nabla_{\gamma, A} \phi = (d+i\Box^ \frac{\gamma-1}{2}A) \Box_  \frac{\gamma-1}{2}\phi.
\eeq
The corresponding curvature will be
\beq
F_{\gamma, A} = (d+i\Box_ \frac{\gamma-1}{2}A) D_{\gamma, A} \phi.
\eeq
Then the equivariance condition 
\beq
D_{\gamma, A} \left( e^{i\Lambda } \odot \phi\right)=  e^{i\Box_\frac{\gamma-1}{2}  \Lambda} D_{\alpha,\beta, A+d_\gamma \Lambda}\phi
\eeq
holds.
\end{theorem}
The proof here is a straightforward generalization of the equivariance condition presented in Sec. 4 of \cite{varna}. For the more general case of a multiform $\Omega$ which is not comprised of $1$-forms, it is easy to write the infinitesimal action of the Gauge group. In case $\Omega=A_0\,dt+ A_{ij} dx^i\wedge dx^j$ the infinitesimal action is given by
\begin{equation}
    A_0\to A_0 + \frac{\partial \Lambda }{\partial t}\qquad A_{ij} \to A_{ij} - \partial _i \partial _j \Lambda
\end{equation}

Writing the global (not infinitesimal) action in general is more complicated. One difficulty we mention is the fact that if $\Omega$ is even just a $p$-form with $p\neq 1$, there is no construction of a $G$-principal bundle of which $\Omega$ is the connection (or no covariant derivative obviously associated with it).  We leave this for later considerations.

We now specialize to the context where $\Omega$ is in fact a mixed tensor $\Omega= A_0 dt +  A_{ij}dx^i \wedge dx^j$.   We then use the transformation
\beq
 \partial_t \rightarrow \partial_t-A_t\,\quad \partial_i \partial_j \rightarrow \partial_i \partial_j -(- \Delta) ^{\frac{\gamma -1}{2}} A_{ij}
\eeq
and the coupling 
\begin{equation}
    \sim\int  \left( (- \Delta) ^{\frac{\gamma -1}{2}} A_{ij}\right) J^{ij}= \int A_{ij}  \,(- \Delta) ^{\frac{\gamma -1}{2}} J^{ij}
\end{equation}
having integrated by parts so that $A_{ij}$ couples with $ (- \Delta) ^{\frac{\gamma -1}{2}} J^{ij}$. 

Recall that the correlation functions of $\rho$ and $J$ are generated by
\beq 
Z[A_t, A_{ij}]= \int exp \left[\,i \int d^{d+1} x \left( A_t \rho +(- \Delta) ^{\frac{\gamma -1}{2}} A_{ij} J^{ij}\right)\, \right]
\eeq 
and the local symmetry implies that under a Gauge transformation $Z[A_t , A_{ij}]=  Z[A_t+ \partial _t \Lambda, A_{ij} -\partial_i \partial_j (- \Delta) ^{\frac{\gamma -1}{2}} \Lambda]$ .

Therefore, by Noether's first theorem
\begin{equation}
    \partial_t \rho +\partial_i \partial_j  \left( (- \Delta) ^{\frac{\gamma -1}{2}} \right) J^{ij}=0,
\end{equation}
meaning the ''current'' is
\begin{equation}
    \hat{J}^i= \partial_j \left( (- \Delta) ^{\frac{\gamma -1}{2}} J^{ij}\right).
\end{equation}
Given that no equation for the current of the form 
\beq    J_{ij}= f(\rho) \,\delta_{ij}
\eeq
is valid as it is inconsistent\cite{gl1,csforms} with $U(1)$ symmetry,  the next  term is
\beq
    J_{ij}= \mathfrak{D}\, \partial_i \partial_j \,\rho \label{n1}
\eeq 
or equivalently 
 \beq
  (- \Delta) ^{\frac{\gamma -1}{2}}  J_{ij}= \mathfrak{D} \,\partial_i \partial_j  (- \Delta) ^{\frac{\gamma -1}{2}}\rho.
\eeq 
Eq.\eqref{n1} is a direct generalization of the standard Fick's law.  From this we infer 
\beq
    \partial_t \,\rho + \mathfrak{D}\,(-\Delta) ^\frac{\gamma +3}{2} \rho=0
\eeq
and therefore we find the associated diffusive  mode,
\begin{equation}
    \omega\,=\,-\,i\,\mathfrak{D}\,k^{\gamma +3}\,,
\end{equation}
which for $\gamma=-1$ reduces to the standard diffusion equation corresponding to the conservation of the total charge $\mathcal{Q}$ associated to the standard $U(1)$ invariance.  For the fracton case, we set $\gamma=1$ and we obtain the result in Ref.\cite{Gromov:2020yoc}. We see then that N\"other's Second Theorem entails a wide variety of modified diffusive behaviour indicative of anomalous diffusion \cite{klafter,fracdiff2016,nigmatullin}.\\

\section{Conserved Charges and Goldstone Effective Action}

Given the most general Lorentz violating Gauge principle, we can now investigate which are the corresponding conserved charges and the effective action for the Goldstone massless modes. Let us start with the example of  subdiffusive dynamics
\begin{equation}
    \langle x^2 \rangle \,\sim\,\mathcal{B}\,t^{1/2},
\end{equation}
which corresponds to the anomalous diffusive mode $\omega=-i \mathfrak{D}k^4$. This dynamics stems from the higher-order constitutive relation
\begin{equation}
   \hat{J}^i= \partial_j J^{ij}\,,\quad  J_{ij}= \mathcal{B} \,\partial_i \partial_j\,\rho
\end{equation}
where $\hat{J}^i$ is the spatial current and $\rho$ the charge density. It is simple to see that this modified constitutive relation follows from the conservation of the electric dipole,
\begin{equation}
    p\,=\,\int d^dx\, \,x\,\rho(x,t) \label{dip}.
\end{equation}
Moreover, it can be derived by gauging the following scalar effective action
\begin{equation}
    L\,=\,c_1 (\partial_t \phi)^2\,-\,c_2\,(\partial_i \partial_j \phi)^2-c_3 (\nabla^2 \phi)^2+\dots \label{act1}
\end{equation}
where terms $\sim (\partial_i \phi)^2$ are absent because they do not conserve the dipole \eqref{dip}. We gauge the action \eqref{act1} by introducing the covariant derivatives
\begin{equation}
    \partial_t \rightarrow \partial_t-A_t\,\quad \partial_i \partial_j \rightarrow \partial_i \partial_j - A_{ij}
\end{equation}
The tensorial current we just discussed comes directly from the minimal coupling with the gauge field, $\sim A_{ij}J^{ij}$. Notice that the conservation of the dipole implies that the dynamics of the emergent quasiparticles is confined along a subspace, in this case a line.

Borrowing from this simple example, we want to find now the general conserved charges and effective actions arising from the non-Lorentz invariant gauge symmetry defined in the previous section. Let us consider a field theory in which the generalized momentum of order $(\gamma+1)/2$ of a certain ''charge'' distribution $\rho(t,x)$ is a conserved quantity,
\begin{equation}
\frac{\partial}{\partial t}\,\mathfrak{m}_{\left(\frac{\gamma+1}{2}\right)}\,=\,\frac{\partial }{\partial t}\,\int\, \sum _i  a_i\, x_i^{\frac{\gamma+1}{2}}\,\rho(t,x)\,dx\,=\,0\,,\label{cont}
\end{equation}
where we remark that if $f(x_i)$ is a function of the sole variable $x_i$, then $(-\Delta )^\gamma f(x_i)= (-\Delta _{x_i})^\gamma f(x_i)$ where $ (-\Delta _{x_i})^\gamma$ is the (fractional) Laplacian in the variable $x_i$.  If $(\gamma+1)/2$ is an integer, then, $\mathfrak{m}_{\left(\frac{\gamma+1}{2}\right)}$ are the standard momenta, e.g. $\gamma=-1$ corresponds to total charge $Q$, $\gamma=1$ dipole $p$, etc. The continuity equation, giving rise to our anomalous diffusive mode, is simply a mathematical re-statement of the principle of conservation in Eq.\eqref{cont}. More precisely, the conservation of $\mathfrak{m}_{\left(\frac{\gamma+1}{2}\right)}$ can be re-written in differential form as the anomalous continuity equation,
\begin{equation}
\partial_t \,\rho + \mathfrak{D}\,(-\Delta) ^\frac{\gamma +3}{2} \rho=0
\end{equation}
In the same way, the effective action for the low-energy degrees of freedom takes the following form,
\begin{equation}
    L\,=\,D_\gamma \left( \star D_\gamma \phi\right)+\dots.
\end{equation}
and it represents a formal generalization of the example shown in Eq.\eqref{act1}.
\section{Conclusions}

It was the purpose of this paper to put under a single umbrella the myriad of diffusive-like equations that have proliferated recently \cite{anomlast,anomlast2} and in the past \cite{tf1,tf2,tf3,klafter,zaslavsky,fracdiffcrit,fracfickslaw,caffdiff} to describe anomalous diffusion.  What we have shown is that NST makes possible a myriad of diffusive-like equations of motion that only reduce to the traditional diffusive limit in special cases.  The source of this multiplicity is the high degeneracy of the eigenvalues of Eq. (\ref{mgen}) which generate a hierarchy of conserved charges.  At play here is the fact that Fick's law is not at all general.  Our work applies equally to Fourier's law for heat transport for which exceptions are well catalogued \cite{fpu}.  The key to this generalization is the non-Lorentzian formulation of the gauge symmetries and hence this work completes the fractional formulation of gauge theories we started previously \cite{csforms,varna,gl1}.

As a result of the work of Caffarelli and Silvestre \cite{cs} who have shown that second-order elliptic differential equations in the upper half-plane in ${\mathbb R}_+^{n+1}$ reduce to one with the fractional Laplacian, $(-\Delta)^\gamma$ at  ${\mathbb R}^n$, where a Dirichlet boundary condition is imposed, our work here has demonstrated that anomalous diffusion can be thought of as a problem in which collective excitations arise in a 
lower-dimensional subspace. While it has been noticed that models requiring higher-rank $U(1)$ gauge theories \cite{PhysRevB.95.115139,PhysRevB.96.035119,PhysRevB.97.235102,Li:2019tje} lead to restricted particle motion in lower dimensional subspaces, the connection to fractional diffusive equations has not been made.  The work presented here suggests that for modes with dispersions (ripplons at superfluid interfaces\cite{mazets,haruki} of the form \begin{equation}
    \omega\,=\,-\,i\,D\,k^{\gamma+3},
\end{equation}
a deep connection exists between the parameter $\gamma$ and the dimensionality of the space for the confined motion in which the effective action is inherently non-local.
\vspace{0.25cm}
\section*{Acknowledgments}
MB thanks W.V.Liu, A.Lucas, A.Gromov, C.Morais-Smith, S.Moroz and D.N.Xuan  for enjoyable discussions and fruitful comments.
M.B. acknowledges the support of the Spanish MINECO ``Centro de Excelencia Severo Ochoa'' Programme under grant SEV-2012-0249.  P.W.P.
thanks DMR19-19143 for partial funding of this project.
\bibliography{hydrobib}
\onecolumngrid
\appendix 
\clearpage
\section*{Appendix: Proof of Theorem III.1}
\begin{proof}
The proof is a simple generalization of the standard fact that holds for the normal Laplacian on forms.  
For simplicity (and without loss of generality) we may assume that $M_1=\mathbb R^{n}$ and $M_2=\mathbb R^{d-n}$ with Euclidean metrics. We then calculate
\begin{equation}
D_\gamma A= \sum _{\substack{k=1\\j_k\in\{ 1, \cdots, n\} }}^{d-p} \frac{\partial}{\partial x^{j_k}}A_{i_1,\cdot , i_p}\, dx^{j_k}\wedge dx^{i_1}\wedge \cdots \wedge dx^{i_p}+\sum _{\substack{k=1\\j_k\in\{ n+1, \cdots, d-n\} }}^{d-p} \frac{\partial}{\partial x^{j_k}}(\Delta _2^{\frac{\gamma-1}{2}} A_{i_1,\cdot , i_p})\, dx^{j_k}\wedge dx^{i_1}\wedge \cdots \wedge dx^{i_p}
\end{equation} whence
\begin{equation}
\star D_\gamma A=\sum _{\substack{k=1\\j_k\in\{ 1, \cdots, n\} }}^{d-p} (-1) ^{p+k-1}\frac{\partial}{\partial x^{j_k}}A_{i_1,\cdot , i_p}\, dx^{j_1}\wedge \cdots \widehat{dx^{j_k}}\wedge  \cdots \wedge dx^{j_{d-p}}+\sum _{\substack{k=1\\j_k\in\{ n+1, \cdots, d-n\} }}^{d-p} (-1) ^{p+k-1}\frac{\partial}{\partial x^{j_k}}(\Delta _2^{\frac{\gamma-1}{2}} A_{i_1,\cdot , i_p})\, dx^{j_1}\wedge \cdots \widehat{dx^{j_k}}\wedge  \cdots \wedge dx^{j_{d-p}}
\end{equation}
which yields
\begin{equation}
\begin{aligned}
&D_\gamma \star D_\gamma A= \sum _{\substack{k=1\\j_k\in\{ 1, \cdots, n\} }}^{d-p} (-1) ^{p+k-1}\frac{\partial^2}{(\partial x^{j_k})^2 }A_{i_1,\cdot , i_p}\,dx^{j_k}\wedge dx^{j_1}\wedge \cdots \widehat{dx^{j_k}}\wedge  \cdots \wedge dx^{j_{d-p}}+\\&\sum _{\substack{k=1\\j_k\in\{ 1, \cdots, n\} }}^{d-p} (-1) ^{p+k-1}\frac{\partial^2}{\partial x^{j_k}\partial x^{i_\ell}}A_{i_1,\cdot , i_p}\,dx^{i_\ell}\wedge dx^{j_1}\wedge \cdots \widehat{dx^{j_k}}\wedge  \cdots \wedge dx^{j_{d-p}}+\\&\sum _{\substack{k=1\\j_k\in\{ n+1, \cdots, d-n\} }}^{d-p} (-1) ^{p+k-1}\frac{\partial^2}{(\partial x^{j_k})^2 }(\Delta _2^{\frac{\gamma-1}{2}} A_{i_1,\cdot , i_p})\, dx^{j_k}\wedge dx^{j_1}\wedge \cdots \widehat{dx^{j_k}}\wedge  \cdots \wedge dx^{j_{d-p}}+\\&\sum _{\substack{k=1\\j_k\in\{ n+1, \cdots, d-n\} }}^{d-p} (-1) ^{p+k-1}\frac{\partial^2}{\partial x^{j_k}\partial x^{i_\ell}}(\Delta _2^{\frac{\gamma-1}{2}} A_{i_1,\cdot , i_p})\, dx^{i_\ell}\wedge dx^{j_1}\wedge \cdots \widehat{dx^{j_k}}\wedge  \cdots \wedge dx^{j_{d-p}}
\end{aligned}
\end{equation}

and also
\begin{equation}
\begin{aligned}
&\star D_\gamma \star D_\gamma A=\\& \sum _{\substack{k=1\\j_k\in\{ 1, \cdots, n\} }}^{d-p} (-1) ^{p+p(d-p)}\frac{\partial^2}{(\partial x^{j_k})^2 }A_{i_1,\cdot , i_p}\,dx^{i_1}\wedge \cdots \wedge dx^{i_p}+\\&\sum _{\substack{k=1\\j_k\in\{ 1, \cdots, n\} }}^{d-p} (-1) ^{pd+\ell}\frac{\partial^2}{\partial x^{j_k}\partial x^{i_\ell}}A_{i_1,\cdot , i_p}\,dx^{j_k} \wedge dx^{1_1}\wedge \cdots \wedge \widehat{ dx^{i_\ell}}\wedge \cdots\wedge dx^{i_p}+\\&\sum _{\substack{k=1\\j_k\in\{ n+1, \cdots, d-n\} }}^{d-p} (-1) ^{p+p(d-p)}\frac{\partial^2}{(\partial x^{j_k})^2 }(\Delta _2^{\frac{\gamma-1}{2}} A_{i_1,\cdot , i_p})\, dx^{i_1}\wedge \cdots \wedge dx^{i_p}+\\&\sum _{\substack{k=1\\j_k\in\{ n+1, \cdots, d-n\} }}^{d-p} (-1) ^{pd+\ell}\frac{\partial^2}{\partial x^{j_k}\partial x^{i_\ell}}(\Delta _2^{\frac{\gamma-1}{2}} A_{i_1,\cdot , i_p})\, dx^{j_k} \wedge dx^{1_1}\wedge \cdots \wedge \widehat{ dx^{i_\ell}}\wedge \cdots\wedge dx^{i_p}.
\end{aligned}
\end{equation} 
A similar calculation shows that 
\begin{equation}
\begin{aligned}
&D_\gamma \star D_\gamma \star A=\\& \sum _{\substack{k=1\\j_k\in\{ 1, \cdots, n\} }}^{d-p} (-1) ^{p(d-p)+d-p+\ell-1}\frac{\partial^2}{(\partial x^{j_k})^2 }A_{i_1,\cdot , i_p}\,dx^{i_1}\wedge \cdots \wedge dx^{i_p}+\\&\sum _{\substack{k=1\\j_k\in\{ 1, \cdots, n\} }}^{d-p} (-1) ^{p(d-p)+d-p+\ell-1}\frac{\partial^2}{\partial x^{j_k}\partial x^{i_\ell}}A_{i_1,\cdot , i_p}\,dx^{j_k} \wedge dx^{1_1}\wedge \cdots \wedge \widehat{ dx^{i_\ell}}\wedge \cdots\wedge dx^{i_p}+\\&\sum _{\substack{k=1\\j_k\in\{ n+1, \cdots, d-n\} }}^{d-p} (-1) ^{p(d-p)+d-p+\ell-1}\frac{\partial^2}{(\partial x^{j_k})^2 }(\Delta _2^{\frac{\gamma-1}{2}} A_{i_1,\cdot , i_p})\, dx^{i_1}\wedge \cdots \wedge dx^{i_p}+\\&\sum _{\substack{k=1\\j_k\in\{ n+1, \cdots, d-n\} }}^{d-p} (-1) ^{p(d-p)+d-p+\ell-1}\frac{\partial^2}{\partial x^{j_k}\partial x^{i_\ell}}(\Delta _2^{\frac{\gamma-1}{2}} A_{i_1,\cdot , i_p})\, dx^{j_k} \wedge dx^{1_1}\wedge \cdots \wedge \widehat{ dx^{i_\ell}}\wedge \cdots\wedge dx^{i_p}
\end{aligned}
\end{equation} and therefore
\begin{equation}
(D_\gamma ^* D_\gamma + D_\gamma D_\gamma ^* ) A= (\Delta _1 A_I )d^I - (\Delta _2 ^\gamma A_I) dx^I
\end{equation}
as we wanted to prove.

\end{proof}
\end{document}